\documentclass{webofc}
\usepackage[utf8]{inputenc}
\usepackage[varg]{txfonts}
\usepackage{booktabs}
\usepackage{xcolor}
\definecolor{darkred}{rgb}{0.4,0.0,0.0}
\definecolor{darkgreen}{rgb}{0.0,0.4,0.0}
\definecolor{darkblue}{rgb}{0.0,0.0,0.4}
\usepackage[bookmarks,linktocpage,colorlinks,
    linkcolor = darkred,
    urlcolor  = darkblue,
    citecolor = darkgreen]{hyperref}
\wocname{EPJ Web of Conferences}

\renewcommand{\d}{\ensuremath{\mathrm d} }
\newcommand{\cD}{\ensuremath{\mathcal D} }
\newcommand{\cO}{\ensuremath{\mathcal O} }
\newcommand{\de}{\ensuremath{\delta} }
\newcommand{\vev}[1]{\ensuremath{\left\langle #1 \right\rangle} }
\newcommand{\eq}[1]{Eq.~\ref{#1}}
\newcommand{\fig}[1]{Fig.~\ref{#1}}
\newcommand{\secref}[1]{Section~\ref{#1}}
\newcommand{\refcite}[1]{Ref.~\cite{#1}}

\begin{document}
\title{Progress applying density of states for gravitational waves}

\author{\firstname{Felix} \lastname{Springer}\inst{1}\fnsep\thanks{Speaker, \email{felix.springer@liverpool.ac.uk}} \and
        \firstname{David} \lastname{Schaich}\inst{1}\fnsep\thanks{\email{david.schaich@liverpool.ac.uk}},
        for the Lattice Strong Dynamics Collaboration} 

\institute{Department of Mathematical Sciences, University of Liverpool, Liverpool L69 7ZL, United Kingdom}

\abstract{%
  Many models of composite dark matter feature a first-order confinement transition in the early Universe, which would produce a stochastic background of gravitational waves that will be searched for by future gravitational-wave observatories. We present work in progress using lattice field theory to predict the properties of such first-order transitions. Targeting SU($N$) Yang--Mills theories, this work employs the Logarithmic Linear Relaxation (LLR) density of states algorithm to avoid super-critical slowing down at the transition.
}
\maketitle

\section{Introduction}
\label{intro}
Standard Markov-chain Monte Carlo importance sampling techniques have been the workhorse for many applications in modern theoretical physics, yet they have some drastic intrinsic shortcomings for certain situations. These include lattice field theory studies of first-order phase transitions, where super-critical slowing down characterized by uncontrollable autocorrelations can result from the difficulty of tunnelling between the two coexisting phases on large lattice volumes~\cite{Borsanyi:2022xml, Langfeld:2022uda}.
In such situations alternatives, such as density of states approaches, may perform better.

First-order phase transitions are currently attracting a great deal of interest due to the possibility that such phenomena in the early Universe could produce an observable background of gravitational waves --- see \refcite{Caprini:2019egz} and references therein.
In this work we focus on gravitational waves from first-order confinement transitions in composite dark matter models, which has also been considered by Refs.~\cite{LatticeStrongDynamics:2020jwi, Huang:2020mso, Kang:2021epo}.
In particular, we consider the Stealth Dark Matter model proposed by the Lattice Strong Dynamics Collaboration~\cite{Appelquist:2015yfa, Appelquist:2015zfa, LatticeStrongDynamics:2020jwi}. This is an SU(4) gauge theory coupled to four fermions that transform in the fundamental representation of the gauge group. While these fundamental fermions are electrically charged, they confine to produce a spin-zero, electroweak-singlet `dark baryon'.

In addition to guaranteeing the stability of a massive dark matter candidate through an analogue of baryon number conservation, the symmetries of Stealth Dark Matter strongly suppress its scattering cross section in direct-detection experiments~\cite{Appelquist:2015zfa, Appelquist:2015yfa}, especially for heavy dark matter masses $M_{\text{DM}} \gtrsim 1$~TeV.
Collider searches for Stealth Dark Matter also become challenging for such heavy masses~\cite{Kribs:2018ilo, Butterworth:2021jto}, which helps to motivate the ongoing work~\cite{LatticeStrongDynamics:2020jwi, Huang:2020mso, Kang:2021epo} investigating how models of this sort could be constrained or discovered by future gravitational-wave observatories such as LISA~\cite{Caprini:2019egz}.

In this proceedings we summarize the progress of our ongoing work applying the Logarithmic Linear Relaxation (LLR) density of states algorithm~\cite{Langfeld:2012ah, Langfeld:2015fua} to analyze SU(4) Yang--Mills theory.
This pure-gauge theory can be considered the `quenched' limit of Stealth Dark Matter corresponding to infinitely heavy fermions.
It is a convenient context in which to consider the LLR algorithm, which is challenging to apply to systems with dynamical fermions~\cite{Korner:2020vjw}.
The SU($N$) Yang--Mills confinement transition is first order for $N \geq 3$, with significantly stronger transitions for $N \geq 4$~\cite{Lucini:2012gg}.
This makes the SU(4) theory a promising first target for application of the LLR algorithm, which could form the basis for future studies of either the SU(3) case relevant for QCD, or of $N \geq 5$ to explore the large-$N$ limit.

In the next section we begin by summarizing the LLR algorithm.
We then discuss in \secref{sec-U(1)} how we tested our LLR code by reproducing some results from \refcite{Langfeld:2015fua} for compact U(1) lattice gauge theory.
In \secref{sec-SU(4)} we present current results from our ongoing LLR analyses of the SU(4) Yang--Mills confinement transition, updating \refcite{Springer:2021liy}.
We wrap up in \secref{sec-outlook} with a brief overview of our planned next steps.

\section{Linear Logarithmic Relaxation}
\label{sec-LLR}
Observables of SU($N$) Yang--Mills theories on the lattice are given by
\begin{align}
  \label{eq:obs}
  \vev{\cO} & = \frac{1}{Z} \int \cD\phi \, \cO(\phi) \, e^{-S[\phi]} &
  Z & = \int \cD \phi \, e^{-S[\phi]},
\end{align}
where $S[\phi]$ is the lattice action and $\phi$ is the set of field variables attached to each link in the lattice. Analytically evaluating these extremely high-dimensional path integrals is practically impossible for the systems we're interested in. Standard Monte Carlo techniques instead aim to give an approximation, by only using a modest number of configurations $\phi$, sampled with probability $\propto e^{-S[\phi]}$.

An alternative approach is to calculate the density of states
\begin{equation}
  \rho(E) = \int \cD \phi \, \delta(S[\phi] - E)
\end{equation}
and use it to reconstruct the observables $\cO$ in \eq{eq:obs} by performing a simple one-dimensional integration over the energy,
\begin{align}
  \vev{\cO(\beta)} & = \frac{1}{Z(\beta)} \int \d E \, \cO(E) \, \rho(E) \, e^{\beta E} &
  Z(\beta) & = \int \d E \, \rho(E) \, e^{\beta E}.
\end{align}
This quantity $\rho(E)$ is not easily accessible.
To calculate it we use the LLR algorithm~\cite{Langfeld:2012ah, Langfeld:2015fua}, which begins by defining the reweighted expectation value
\begin{align}
  \vev{\vev{E - E_i}}_{\de}(a) & = \frac{1}{N}\int \cD \phi \, (E-E_i) \, \theta_{E_i,\de} \, e^{-aS[\phi]} = \frac{1}{N} \int_{E_i-\frac{\de}{2}}^{E_i+\frac{\de}{2}} \d E \, (E-E_i) \, \rho(E) \, e^{-aE}, \label{Heavyside} \\
  N & = \int \cD \phi \, \theta_{E_i,\de} \, e^{-aS[\phi]} = \int_{E_i-\frac{\de}{2}}^{E_i+\frac{\de}{2}} \d E \, \rho(E) \, e^{-a E}.
\end{align}
Here $E_i$ is a fixed energy value, $\theta_{E_i,\de}$ is the modified Heaviside function ($1$ in the interval $E_i \pm \frac{\delta}{2}$ and $0$ everywhere else), and for now `$a$' is just a free parameter not to be confused with the lattice spacing.

The next step is to set the reweighted expectation value $\vev{\vev{E - E_i}}_{\de}(a)$ to zero and approximate the integral with the trapezium rule:
\begin{align}
  \vev{\vev{E - E_i}}_{\de}(a) & = \frac{1}{N} \int_{E_i-\frac{\de}{2}}^{E_i+\frac{\de}{2}} \d E \, (E-E_i) \, \rho(E) \, e^{-aE} \\
  & =\frac{1}{N} \frac{\de}{2}\left( (\frac{\de}{2})e^{-a(E_i+\frac{\delta}{2})}\rho(E_i+\frac{\de}{2})+(-\frac{\de}{2})e^{-a(E_i-\frac{\de}{2})}\rho(E_i-\frac{\de}{2})\right) + \cO(\de^3) = 0. \nonumber
\end{align}
After expanding the exponential $e^{\pm a \frac{\de}{2}}$ and the density $\rho(E_i\pm \de/2)$ in a Taylor series and neglecting $\cO(\de^2)$ terms in the limit $\de \rightarrow 0$, we arrive at the following expression for the parameter $a$:
\begin{align}
  0 & = \left(\rho(E_i) + \frac{\de}{2} \frac{\d \rho(E)}{\mathrm{d}E}\Bigr|_{E=E_i}\right)\left(1-a\frac{\de}{2}\right) - \left(\rho(E_i)-\frac{\de}{2}\frac{\d \rho(E)}{\d E}\Bigr|_{E=E_i}\right)\left(1+a\frac{\de}{2}\right) \nonumber \\
  & = \left(-\rho(E_i)a + \frac{\d \rho(E)}{\d E}\Bigr|_{E = E_i} - \rho(E_i)a + \frac{\d \rho(E)}{\d E}\Bigr|_{E=E_i}\right)\frac{\de}{2} \\
  \implies a & = \frac{1}{\rho(E_i)}\frac{\d \rho(E)}{\d E}\Bigr|_{E=E_i} = \frac{\d \ln(\rho(E))}{\d E}\Bigr|_{E=E_i}. \label{eq:a}
\end{align}
We can see from this, that $a(E_i)$ is a linear approximation of the derivative of the logarithm of the density of states $\rho(E_i)$.
The full density of states $\rho(E)$ can now be reconstructed, with an exponentially suppressed error~\cite{Langfeld:2012ah, Langfeld:2015fua, Langfeld:2016kty}, by performing a numerical integration of the values of $a(E_i)$ across many intervals $E_i \pm \frac{\de}{2}$ and exponentiating the result.

To compute $a$ for a given $E_i$, we use the Robbins--Monro algorithm to solve the equation $\vev{\vev{E - E_i}}_{\de}(a) = 0$~\cite{Langfeld:2015fua}:
\begin{equation}
  a^{(n+1)}=a^{(n)}+ \frac{12}{\de^2 (n+1)}\vev{\vev{E - E_i}}_{\de}(a^{(n)}).
  \label{eq:robmon}
\end{equation}
This series has a fixed point at the correct value of the LLR parameter $a=a^{(n+1)}=a^{(n)}$. At each iteration $j$ we evaluate the reweighted expectation value $\vev{\vev{E - E_i}}_{\de}(a^{(j)})$ using standard importance-sampling Monte Carlo techniques, but with the probability weight $e^{-a^{(j)}S}$ rather than the usual $e^{-S}$.

The modified Heaviside function $\theta_{E_i,\de}$ in \eq{Heavyside} means we reject all Monte Carlo updates that produce a configuration with an energy outside of $E_i \pm \frac{\delta}{2}$, potentially causing low acceptance rates for small energy intervals $\delta$. Alternatively we can substitute this hard energy cut-off with a smooth Gaussian window function~\cite{Langfeld:2016kty, Korner:2020vjw}:
\begin{equation}
  \label{eq:gauss}
  \vev{\vev{E - E_i}}_{\de}(a) = \frac{1}{N} \int \d E \, (E-E_i) \, \exp\left[-\frac{(E - E_i)^2}{2\de^2}\right] \, \rho(E) \, e^{-aE}.
\end{equation}
This effectively changes the probability weight in the Monte Carlo simulation to $\exp\left[-\frac{(E - E_i)^2}{2\de^2}\right] e^{-aE}$. Unlike the modified Heaviside function, the Gaussian window function is differentiable, enabling the use of the hybrid Monte Carlo (HMC) algorithm in the evaluation of the reweighted expectation value.
In our work we test and compare both ways of constraining the energy interval.

\section{U(1) bulk phase transition}
\label{sec-U(1)}
As an initial test of our implementation of the LLR algorithm, we reproduced some results from \refcite{Langfeld:2015fua} for four-dimensional pure-gauge U(1) theory, defined by the action
\begin{equation}
  S = -\beta \sum_{x,\mu < \nu} \cos(\theta_{\mu \nu}(x)),
\end{equation}
with $\theta_{\mu \nu}(x) = \theta_{\mu}(x) + \theta_{\nu}(x + \hat{\mu}) - \theta_{\mu}(x+\hat{\nu}) - \theta_{\nu}(x)$.  The compact link variable $\theta_{\mu}(x) \in [-\pi; \pi]$ is attached to the lattice site $x$ in direction $\hat{\mu}$ and the sum runs over each link in the lattice. $\beta = \frac{1}{g_0^2}$ with $g_0^2$ being the bare gauge coupling.

\begin{figure}[btp]
  \includegraphics[width=0.49\textwidth]{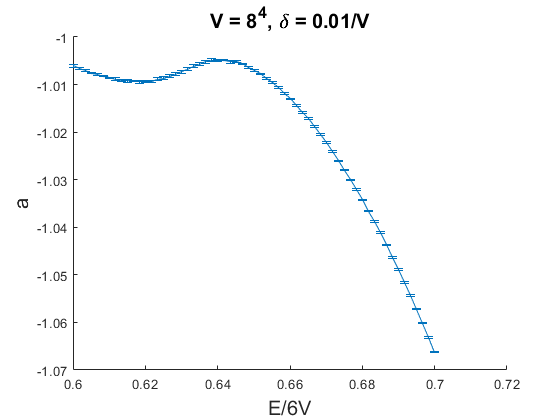}\hfill
  \includegraphics[width=0.49\textwidth]{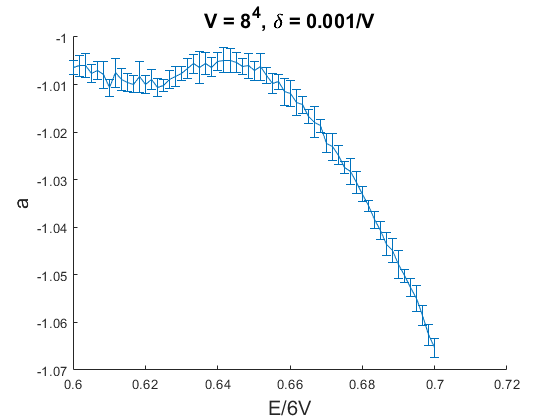}
  \caption{Results for $a$ obtained for compact U(1) pure-gauge theory with lattice volume $V=8^4$ and energy intervals of size $\de = 0.01/V$ (left) or $\de = 0.001/V$ (right).  We can see that the statistical uncertainties increase in the limit $\de \to 0$ considered in \eq{eq:a}.} 
  \label{fig:a_U1}
\end{figure}

For this simple system, we use the Metropolis--Rosenbluth--Teller algorithm with randomly updated angles $\theta_{\mu}(x)$, imposing a hard energy cut-off (\eq{Heavyside}) to evaluate the reweighted expectation value $\vev{\vev{E - E_i}}_{\de}(a^{(n)})$. 
As shown in \refcite{Springer:2021liy}, our results for the values of $a$ are consistent with the results from \refcite{Langfeld:2015fua}, providing confidence in our underlying implementation of the LLR algorithm.
Figure~\ref{fig:a_U1} shows our results for $a$, using two different values of the energy interval size, $\de = 0.01/V$ and $0.001/V$.
While we are interested in the limit $\de \to 0$ from \eq{eq:a}, we can see that smaller \de results in larger statistical uncertainties.
In both plots the error bars are determined through a jackknife analysis of $N_{\text{j}}=10$ completely independent calculations for each energy interval.

\begin{figure}[btp]
  \includegraphics[width=0.49\textwidth]{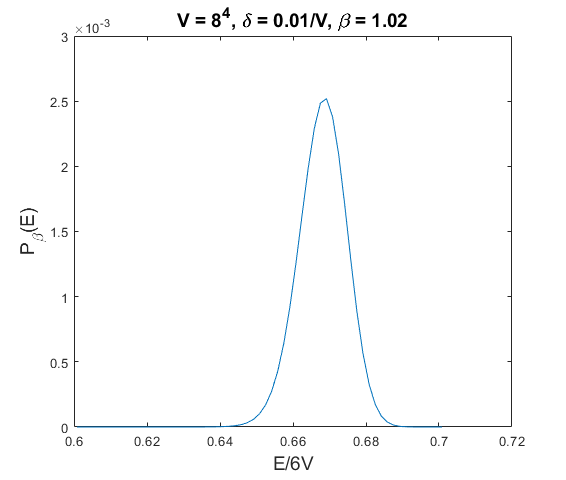}\hfill
  \includegraphics[width=0.49\textwidth]{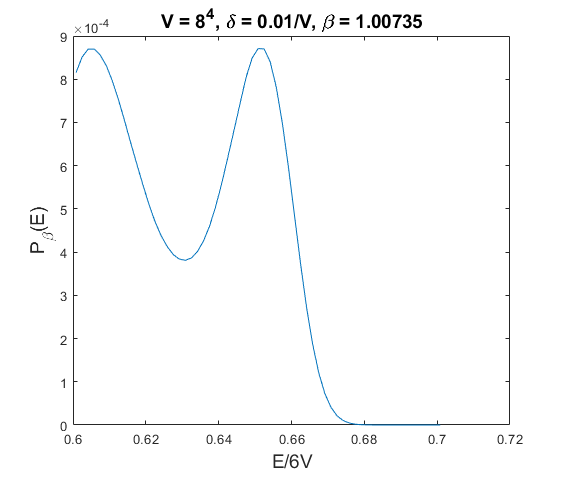}
  \caption{Comparison between the probability densities $P_{\beta}(E) = \rho(E) e^{\beta E}$ (omitting uncertainties) for compact U(1) lattice gauge theory away from the phase transition ($\beta=1.02$, left) and at the phase transition ($\beta_c=1.00735$, right), using lattice volume $V=8^4$ and energy interval size $\delta = 0.01/V$.}
  \label{fig:rho_U1}
\end{figure}

From these results for $a$ we can reconstruct the probability density $P_{\beta}(E) = \rho(E) e^{\beta E}$ via a simple trapezium-rule numerical integration. Figure~\ref{fig:rho_U1} illustrates the sensitivity of this probability density to the value of $\beta$. While $P_{\beta}(E)$ has only a single peak for $\beta = 1.02$, it has a clear double-peak structure for $\beta_c = 1.00735$, signalling the coexistence of the two phases at the first-order bulk phase transition. The latent heat $\Delta E / (6V) \approx 0.045$ of this first-order phase transition can be directly read off as the separation between the two peaks.

\section{SU(4) phase transition}
\label{sec-SU(4)}
We now turn to our ongoing work applying the LLR algorithm to SU(4) Yang--Mills theory. The action is
\begin{equation}
  \label{eq:wilson}
  S = -\beta \sum_{x,\mu<\nu} \mathrm{Re}\mathrm{Tr}\left(U_{\mu\nu}(x)\right),
\end{equation}
with the plaquette $U_{\mu\nu}(x) = U_{\mu}(x) U_{\nu}(x+\hat{\mu}) U_{\mu}^{\dagger}(x+\hat{\nu}) U_{\nu}^{\dagger}(x)$.
Here $\beta = \frac{8}{g_0^2}$, with $g_0^2$ the bare gauge coupling, the sum runs over all lattice sites and $U_{\mu}(x)$ is the SU(4)-valued link variable attached to lattice site $x$ in direction $\hat{\mu}$.

We implemented the LLR algorithm in a fork of Stefano Piemonte's \texttt{LeonardYM} software.\footnote{\texttt{\href{https://github.com/FelixSpr/LeonardYM}{github.com/FelixSpr/LeonardYM}}}
For the calculation of the reweighted expectation value $\vev{\vev{E - E_i}}_{\de}(a^{(n)})$, we have compared three different updating schemes: overrelaxation updates in the full SU($N$) group~\cite{Creutz:1987xi}, the Metropolis--Rosenbluth--Teller algorithm with SU($N$) updates generalized from \refcite{Katznelson:1984kw}, and HMC updates.
While our use of the HMC algorithm requires the smooth Gaussian window function \eq{eq:gauss}, in the other two cases we have further compared both this Gaussian window approach and the hard energy cut-off of \eq{Heavyside}.
We obtain consistent results from all of these updating schemes.

\begin{figure}[btp]
  \centering
  \includegraphics[width=6.5cm]{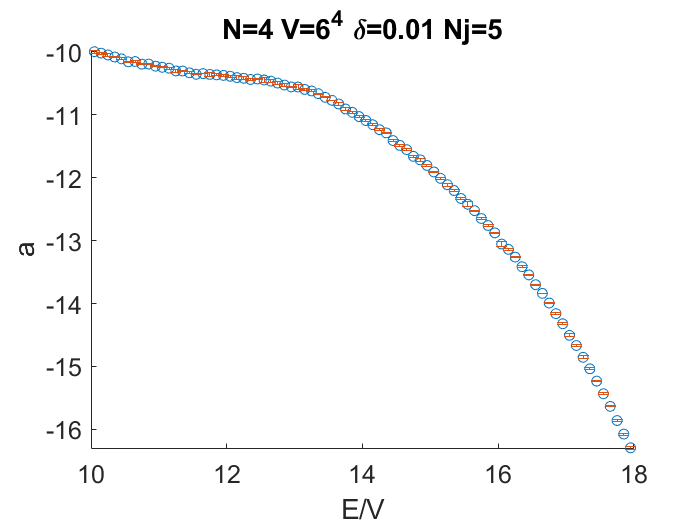}\hfill
  \caption{SU(4) results for $a$ from $6^4$ lattices with an energy interval size of $\de = 0.01/V$.  The statistical uncertaintiess are obtained by performing $N_{\text{j}} = 5$ independent runs per interval.}
  \label{fig:avse}
\end{figure}

For SU($N$) Yang--Mills theories on an $N_s^3 \times N_t$ lattice, there are two distinct transitions we could consider.
The physically relevant confinement transition is first order for $N \geq 3$ (but only weakly so for $N = 3$)~\cite{Lucini:2012gg}.
This transition occurs at a fixed temperature $T_c = 1 / (a N_t)$ (with this $a$ the lattice spacing), implying $\beta_c \to \infty$ as $N_t \to \infty$.
In addition, there is a bulk transition at an $N_t$-independent coupling $\beta_{\text{bulk}}$, which is first order for $N \geq 5$ (but only weakly so for $N = 5$)~\cite{Lucini:2012gg}.
If $N_t$ is too small, $\beta_c \approx \beta_{\text{bulk}}$ and the confinement transition is distorted by the nearby bulk transition --- even for $N \leq 4$ where the latter is a continuous crossover.
Based on prior work including Refs.~\cite{Wingate:2000bb, LatticeStrongDynamics:2020jwi}, we consider $N_t \geq 6$ in order to avoid this problem.

As a brief aside on the bulk transition, in \fig{fig:avse} we show results for the LLR parameter $a$ across a wide range of energies for $6^4$ lattices.
Although we can see that the decrease of $a$ decelerates around the energy of the bulk crossover, the function $a(E)$ remains monotonic in contrast to \fig{fig:a_U1} for the U(1) case. As a result the reconstructed probability density $P_{\beta}(E)$ does not feature a double-peak structure for any value of $\beta$, confirming a continuous crossover with no phase coexistence.
In ongoing work soon to be reported, we will directly contrast this with the first-order bulk transition of SU(6) Yang--Mills theory.

\begin{figure}[btp]
  \centering
  \includegraphics[width=6.5cm]{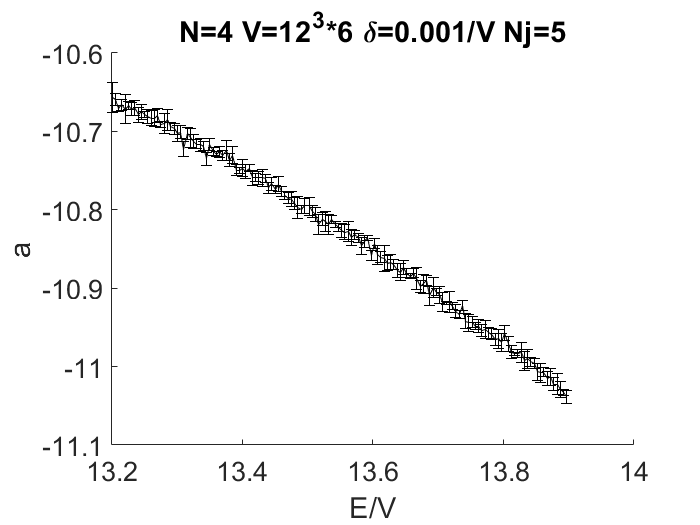}\hfill
  \caption{SU(4) results for $a$ from $V=12^3 \times 6$ lattices with an energy interval size of $\delta = 0.001/V$.  The error bars are estimated by performing $N_{\text{j}} = 5$ independent runs per interval. There is no sign of a first-order phase transition, which would correspond to a non-monotonic $a(E)$.}
  \label{fig:avse_conf}
\end{figure}

To investigate the physically interesting first-order confinement transition of pure-gauge SU(4), we have to use a lattice with an aspect ratio $r \equiv N_s / N_t > 1$.
While we have carried out LLR calculations using a range of lattice volumes, here we focus on $V=12^3 \times 6$, also fixing the energy interval size $\de = 0.001/V$ and $N_{\text{j}}=5$ independent runs of the Robbins--Monro algorithm for each energy interval.
Based on $N_t = 6$ importance sampling results in \refcite{Wingate:2000bb}, in \fig{fig:avse_conf} we focus on the energy range $13.2 < E/V < 13.9$.
There is no sign of a non-monotonic $a(E)$, which continues to be the case throughout larger ranges of $E/V$, implying that we are not yet able to resolve the first-order confinement transition.
We have confirmed this by reconstructing the probability density $P_{\beta}(E)$ using both naive trapezium-rule integration and a more robust polynomial fit technique~\cite{Francesconi:2019nph,Francesconi:2019aet}.
As expected, neither method produces a double-peak structure for any $\beta$.

\begin{figure}[btp]
  \includegraphics[width=5.8cm]{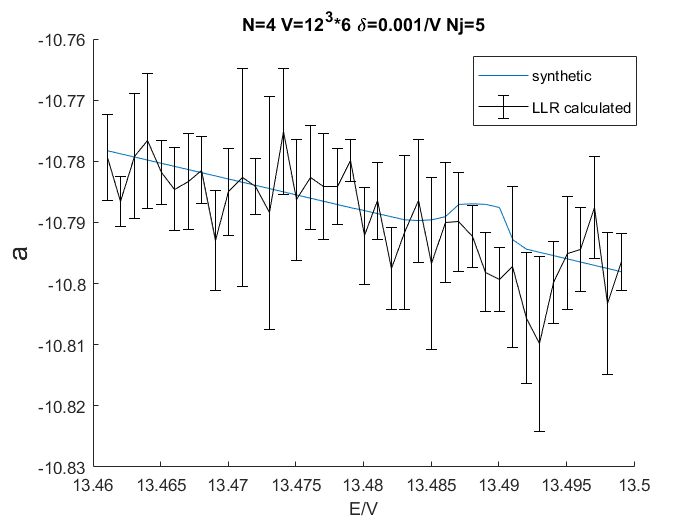}\hfill
  \includegraphics[width=6.0cm]{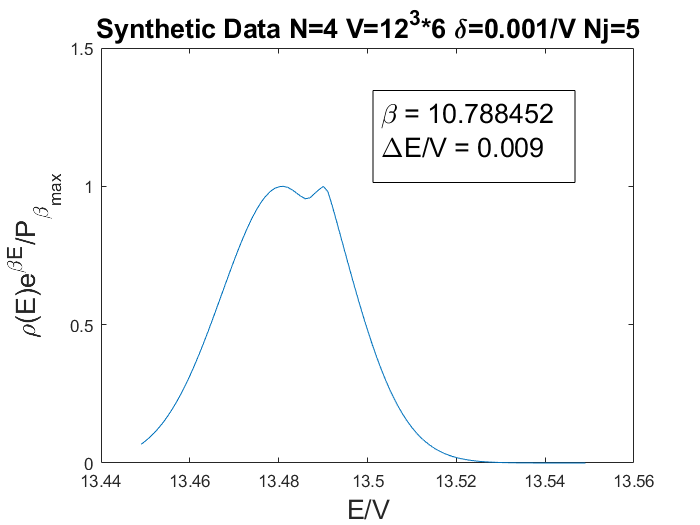}
  \caption{The left plot compares numerical data for $a$ (black) with the synthetically generated values (blue line) that would be needed to reconstruct a similar first-order confinement transition to the one in \refcite{Wingate:2000bb}.  Both the synthetic and the measured $a$ use lattice volume $V=12^3 \times 6$.  The right plot shows the probability density $P_{\beta}(E) = \rho(E) e^{\beta E}$ reconstructed from the synthetic $a(E)$ with $\beta \approx 10.8$. A small double-peak structure is visible, confirming a first-order transition.}
  \label{fig:comparison_synth}
\end{figure}

We can gain some appreciation for why we have not yet observed a first-order SU(4) confinement transition by noting that the $N_t = 6$ importance-sampling results in \refcite{Wingate:2000bb} imply a latent heat of $\Delta E / V \approx 0.004$ at $\beta_c \approx 10.8$.
This latent heat is roughly two orders of magnitude smaller than that in the U(1) case, so it is not necessarily surprising that our LLR calculations have not yet resolved it.
To test this hypothesis, we generated synthetic values of $a$ that would correspond to a first-order phase transition with a similar pseudo-critical $\beta_c \approx 10.8$ and latent heat $\Delta E / V \approx 0.004$.
In \fig{fig:comparison_synth} we overlay our numerical data on top of this synthetic $a(E)$ (blue line), which allows us to conclude that our statistical precision is currently insufficient to resolve the non-monotonocity corresponding to such a first-order phase transition.
These considerations provide some clear next steps for our work, which we discuss in the next section.

\section{Outlook and next steps}
\label{sec-outlook}
In this proceedings we have presented our progress applying the LLR density of states algorithm to investigate first-order transitions in lattice gauge theories.
By analyzing the density of states we aim to avoid the super-critical slowing down that plague Markov-chain importance-sampling techniques at such first-order transitions.
Motivated by the Stealth Dark Matter model and the stochastic gravitational-wave background that would be produced by a first-order dark-sector confinement transition in the early Universe, our ongoing work focuses on SU(4) Yang--Mills theory.

Our current results from $12^3 \times 6$ lattices are insufficient to resolve the SU(4) confinement transition.
While a smaller energy interval size \de could help, given the small latent heat of this transition, this will also make it more challenging to control statistical uncertainties as shown by \fig{fig:a_U1}.
We are currently improving our analyses by analyzing larger $N_s^3 \times N_t$ lattice volumes, both increasing $N_s$ with fixed $N_t$ to study the thermodynamic limit and increasing $N_t$ with fixed aspect ratio $N_s / N_t$ to extrapolate to the continuum limit.
Once we have resolved the first-order confinement transition, we will be able to predict observables such as the latent heat and surface tension that affect the resulting spectrum of gravitational waves.

From our earlier results for compact U(1) lattice gauge theory, we can appreciate that, in contrast to standard importance sampling approaches, the LLR method may be easiest to apply to study very strong first-order phase transitions characterized by a large latent heat.
Because the first-order SU($N$) bulk transition for $N > 4$ is notably stronger than the finite-temperature transition, we are separately using the LLR algorithm to study such large-$N$ bulk transitions and explore the expected scaling of the latent heat $\propto N^2$~\cite{Lucini:2012gg}.
We will soon present preliminary results for the SU(5) and SU(6) bulk phase transitions.

Based on these studies of both the large-$N$ bulk transition and SU(4) confinement transition, we will be in a good position to consider the confinement transition for SU($N$) Yang--Mills theories with larger $N$. 
We will also be able to estimate the computing resources that would be required to analyze the weaker first-order confinement transition of SU(3) Yang--Mills theory corresponding to quenched QCD.
Finally, we are exploring broader applications of the LLR algorithm, including to phase transitions in bosonic matrix models of interest in the context of holographic gauge/gravity duality.

\vspace{20 pt}
\noindent \textsc{Acknowledgments:}~We thank the LSD Collaboration for ongoing joint work investigating composite dark matter and gravitational waves.
We also thank Kurt Langfeld, Paul Rakow, David Mason, James Roscoe and Johann Ostmeyer for helpful conversations about the LLR algorithm.
Numerical calculations were carried out at the University of Liverpool and through the Lawrence Livermore National Laboratory Institutional Computing Grand Challenge program.
DS was supported by UK Research and Innovation Future Leader Fellowship {MR/S015418/1} and STFC grant {ST/T000988/1}.

\bibliography{main}

\end{document}